\documentclass[12pt,thmsa]{article}
\usepackage{sw20lart}

\input{tcilatex}

\begin{document}

\title{$SU_{L}(4)\times U(1)$ model for electroweak unification and sterile
neutrinos}
\author{Riazuddin \\
National Centre for Physics\\
Quaid-i-Azam University\\
Islamabad, Pakistan. \and Fayyazuddin \\
Department of Physics and\\
National Centre for Physics, Quaid-I-Azam University\\
Islamabad, Pakistan.}
\maketitle

\begin{abstract}
Some of basic problems in neutrino physics such as new energy scales, the
enormous gap between neutrino masses and the lightest charged fermion mass,
possible existance of sterile neutrinos in eV mass range are studied in the
local gauge group $SU_{L}(4)\times U(1)$ for electroweak unification, which
does not contain fermions with exotic electric charges. It is shown that
neutrino mass spectrum can be decoupled from that of the other fermions.
Further normal seesaw mechanism for neutrinos, with right handed neutrino
Majorana masses to be of order $M\gg M_{\text{weak}}$ as well a new eV-scale
can be accomodated. The eV-scale seesaw may manifest itself in experiments
like Liquid Scintilation Neutrino Detector (LSND) and MiniBooNE (MB)
experimental results and future neutrino experiments.
\end{abstract}

In recent years enormous progress has been made in neutrino physics, which
also has relevance to many fields other than the particle physics; in
particular, nuclear physics, astrophysics and cosmology. This has been made
possible by the quantum mechanical phenomena of interferometory which
provides sensitive method to explore extremely small effects. This has
resulted in the discovery of neutrino oscillations which imply that they
have tiny but finite masses against the prediction of the standard model of
particle physics. Thus they provide evidence for new physics which goes
beyond the standard model. New physics requires new energy scale beyond that
provided by the standard model but such a scale has not yet been pinned
down. Thus there is a need to consider extensions of the electro-weak group
which would provide a new scale between the elctroweak and grand
unification. The neutrinos may also provide an understanding of the origin
of matter (baryogenesis) through leptogenesis. For this purpose right handed
neutrinos, which are seesaw partners of light neutrinos, with a mass scale
of $10^{10}-10^{11}$ GeV or even in TeV region may be needed. Further more
one sees the enormous gap between neutrino masses, revealed by neutrino
oscillations, and the lightest charged fermion mass $\left( m_{e}\right) $
in contrast to that between $m_{e}$ and $m_{t}$ (top quark mass), which is
populated by charged leptons and quarks. Further $\left( m_{\nu }\right) _{%
\text{max}}/m_{e}<2\times 10^{-6}$ which needs to be undersood. This may be
an indication of decoupling of neutrino mass spectrum from other fermions.
Morover while all neutrino data can be explained by flavor oscillations of
three active neutrinos \cite{1a}, the Liquid Scintilation Neutrino Detector
(LSND) anomaly \cite{1} stands out. This anomaly together with MiniBooNE
(MB) experiment \cite{2} may require atleast two sterile neutrinos \cite{3},%
\cite{4} that mix with the active neutrinos. Another possibility is by a
decaying sterile neutrino, again in eV range \cite{5}. Their mass is in the
range of electrovolts. The purpose of this paper is a modest attempt towards
understanding of some the problems mentioned above.

We consider the extension of electroweak group $SU_{L}\left( 2\right) \times
U\left( 1\right) $ to $SU_{L}(4)\times U_{X}(1)$ as such an extension can
answer some of the questions raised above as we shall see. In particular, it
is shown that in addition to normal seesaw mechanism for neutrino masses,
where right handed Majorana mass$\gg M_{weak}$ a new eV-scale can be
accomodated. The latter may be a manifestation of LSND and MB experiments.
Further we show that neutrino mass spectrum can be decoupled from that of
charged leptons.

If one restricts oneself to only $SU_{L}(4)$\cite{6}, one can not only
accomodate known leptons nicely, but also a right handed Majorana neutrino.
However in order to accomodate the quarks, the group has to be extended to $%
SU_{L}(4)\times U_{X}(1)$\cite{7}, \cite{8},\cite{9}. In the original
minimal version, where leptons $(l^{c},\nu _{l}^{c},N_{l},l)_{R}$ form an $%
SU(4)$ quartet, in order to cancel the anomalies, one has to have quarks
with exotic electric charges $\frac{4}{3}$ and $\frac{5}{3}$. In this paper
we shall consider other versions, which do not involve quarks with exotic
charges\cite{10}.

The electric charge operator can, in general be defined as a linear
combination of diagonal generators of the group [$\widehat{I}$, being the
unit matrix] 
\begin{eqnarray}
Q &=&\frac{1}{2}[\lambda _{3}+\frac{b}{\sqrt{3}}\lambda _{8}+\frac{2c}{\sqrt{%
6}}\lambda _{15}]+\frac{Y_{X}}{2}\widehat{I}  \nonumber \\
&=&diag[\frac{1}{2}+\frac{b}{6}+\frac{c}{6}+\frac{Y_{X}}{2},\text{ }-\frac{1%
}{2}+\frac{b}{6}+\frac{c}{6}+\frac{Y_{X}}{2},\text{ }-\frac{2b}{6}+\frac{c}{6%
}+\frac{Y_{X}}{2},\text{ }-\frac{c}{2}+\frac{Y_{X}}{2}]  \nonumber \\
&=&\frac{1}{2}(\lambda _{3}+\widehat{Y}_{1})+\frac{Y_{X}}{2}\widehat{I}
\label{1}
\end{eqnarray}
where $Y_{X}$ is the hypercharge associated with $U_{X}$ and 
\begin{equation}
\widehat{Y}_{1}=diag[\frac{b+c}{3},\text{ }\frac{b+c}{3},\text{ }\frac{-2b+c%
}{3},\text{ }-c]  \label{2}
\end{equation}
Now 
\begin{eqnarray*}
\frac{1}{e^{2}} &=&\frac{1}{g^{2}}+\frac{1}{g^{\prime 2}} \\
\frac{1}{g^{\prime 2}} &=&\frac{1}{g_{1}^{2}}+\frac{1}{g_{X}^{2}}
\end{eqnarray*}
where in the $SU_{L}(4)$ limit 
\begin{eqnarray}
g_{1}^{2} &=&\frac{1}{C_{1}^{2}}g^{2}  \nonumber \\
C_{1}^{2} &=&\frac{b^{2}+2c^{2}}{3}  \label{3}
\end{eqnarray}
This gives 
\[
\frac{1}{g^{\prime 2}}=\frac{b^{2}+2c}{3g^{2}}+\frac{1}{g_{X}^{2}} 
\]
Since $\frac{g^{\prime }}{g}=\tan \theta _{W}$, one obtains 
\begin{equation}
\frac{g_{X}^{2}}{g^{2}}=\frac{\sin ^{2}\theta _{W}(m_{X})}{1-\frac{%
3+b^{2}+2c^{2}}{3}\sin ^{2}\theta _{W}(m_{X})}  \label{4}
\end{equation}
In the minimal version, $b=1,$ $c=2$%
\begin{equation}
\frac{g_{X}^{2}}{g^{2}}=\frac{\sin ^{2}\theta _{W}(m_{X})}{1-4\sin
^{2}\theta _{W}(m_{X})}  \label{5}
\end{equation}
and $Q=(1,0,0,-1)+\frac{Y_{X}}{2}(1,1,1,1),$ so that for leptons, $Y_{X}=0,$%
where as in order to accomodate quarks we take, $Y_{X}=-\frac{2}{3}$ for the
first two generations of quarks and $Y_{X}=-\frac{1}{3}$ for the third
generation. This is because in order to cancel the anomalies, one generation
is to be treated differently from the other two. In this case we have quarks
with exotic electric charges $-\frac{4}{3}$ and $-\frac{5}{3}$ respectively.

In order to accomodate known isospin doublets of left-handed quarks and
leptons in the two upper components of $4$ and $4^{*}$(or $4^{*},$ 4)
representations of $SU(4)$ and to forbid exotic electrical charges, we must
have $\frac{b+c}{6}=\pm \frac{1}{4},$ $\frac{-2b+c}{6}=-\frac{c}{2}$ so that 
$b=2c=\pm 1.$ We thus consider the version with $b=2c=1$ (the other choice
is equivalent). This choice gives [$C_{1}^{2}=\frac{1}{2}$] 
\begin{equation}
\frac{g_{X}^{2}}{g^{2}}=\frac{\sin ^{2}\theta _{W}(m_{X})}{1-\frac{3}{2}\sin
^{2}\theta _{W}(m_{X})}  \label{6}
\end{equation}
A straight forward application of renormalization group equations gives\cite%
{7} [$\sin ^{2}\theta _{W}=\sin ^{2}\theta _{W}(m_{Z})$] 
\begin{eqnarray*}
&&1-(1+C_{1}^{2})\sin ^{2}\theta _{W}-\frac{\alpha _{X}^{-1}(m_{Z})}{\alpha
^{-1}(m_{Z})} \\
&=&2\frac{\alpha (m_{Z})}{4\pi }[-C_{1}^{2}(-\frac{22}{3}+\frac{4}{3}\frac{nf%
}{2})+\frac{4}{3}\frac{nf}{2}C_{1}^{2}]\ln \frac{m_{X}}{m_{Z}} \\
&=&\frac{\alpha (m_{Z})}{4\pi }\frac{44}{3}C_{1}^{2}\ln \frac{m_{X}}{m_{Z}}
\end{eqnarray*}
For our case $C_{1}^{2}=\frac{1}{2}$ and we obtain 
\[
1-\frac{3}{2}\sin ^{2}\theta _{W}-\frac{\alpha (m_{Z})}{\alpha _{X}(m_{Z})}=%
\frac{22}{3}\frac{\alpha (m_{Z})}{4\pi }\ln \frac{m_{X}}{m_{Z}} 
\]
where $\sin ^{2}\theta _{W}(m_{Z})=0.23122$ and $\alpha ^{-1}(m_{Z})=128.$
The unification scale $m_{X}$ is not very sensitive to $\alpha _{X}.$For $%
m_{X}=10^{3}$GeV, $10^{6}$GeV, $10^{10}$GeV, $10^{16}$GeV, $\alpha
_{X}=1.22\times 10^{-2}$, $1.28\times 10^{-2},$ $1.37\times 10^{-2},$ $%
1.54\times 10^{-2}$ respectively. To put it in proper perspective, we note
that the coupling $\alpha ^{\prime }=\frac{\alpha }{\cos ^{2}\theta _{w}}$
associated with $U(1)$ of the Standard Model is $\simeq 1.30\alpha \simeq
1.02\times 10^{-2}.$Before we give the anomaly free fermion content, we note
that for our choice $b=2c=1,$%
\[
Q=diag[\frac{3}{4}+\frac{Y_{X}}{2},-\frac{1}{4}+\frac{Y_{X}}{2},-\frac{1}{4}+%
\frac{Y_{X}}{2},-\frac{1}{4}+\frac{Y_{X}}{2}] 
\]
We can have two possibilities; which are given in Table 1 below. 
\begin{eqnarray*}
&& 
\begin{tabular}{|l|}
\hline
TableI \\ \hline
\end{tabular}
\text{Anomaly free fermion content(}i\text{ is the generation index,} \\
&&\text{ }a\text{ is the color index and }c\text{ stands for charge
conjugation)} \\
&& 
\begin{tabular}{|l|l|}
\hline
$SU(4)quartet:$ & $SU(4)$ singlet: \\ \hline
$I$ &  \\ \hline
$F_{iL}^{l}=\left( 
\begin{array}{l}
\nu _{i} \\ 
e_{i}^{-} \\ 
E_{i}^{-} \\ 
F_{i}^{-}%
\end{array}
\right) _{L,\text{ }Y_{X}=-\frac{3}{2}}i=1,2,3$ & $%
\begin{tabular}{l}
$F_{iR}^{l}\equiv \ \left( 
\begin{array}{llll}
N_{i} & e_{i}^{-} & E_{i}^{-} & F_{i}^{-}%
\end{array}
\right) _{R}$ \\ 
$Y_{X}:\quad \left( 
\begin{array}{llll}
0 & -2 & -2 & -2%
\end{array}
\right) $%
\end{tabular}
$ \\ \hline
$F_{1L}^{q}=\left( 
\begin{array}{l}
u_{1}^{a} \\ 
d_{1}^{a} \\ 
D_{1}^{a} \\ 
H_{1}^{a}%
\end{array}
\right) _{L,\text{ }Y_{X}=-\frac{1}{6}}$ & $%
\begin{tabular}{l}
$F_{1R}^{q}\equiv \ \left( 
\begin{array}{llll}
u_{1}^{a} & d_{1}^{a} & D_{1}^{a} & H_{1}^{a}%
\end{array}
\right) _{R}$ \\ 
$Y_{X}:\quad \left( 
\begin{array}{llll}
\frac{4}{3} & -\frac{2}{3} & -\frac{2}{3} & -\frac{2}{3}%
\end{array}
\right) $%
\end{tabular}
$ \\ \hline
$F_{iR}^{q}=\left( 
\begin{array}{l}
d_{i}^{ac} \\ 
u_{i}^{ac} \\ 
U_{i}^{ac} \\ 
T_{i}^{ac}%
\end{array}
\right) _{R,\text{ }Y_{X}=-\frac{5}{6}}i=2,3$ & $%
\begin{tabular}{l}
$F_{iR}^{q}\equiv \ \left( 
\begin{array}{llll}
d_{i}^{ac} & u_{i}^{ac} & U_{i}^{ac} & T_{i}^{ac}%
\end{array}
\right) _{R}$ \\ 
$Y_{X}:\quad \text{ }\left( 
\begin{array}{llll}
\frac{2}{3} & -\frac{4}{3} & -\frac{4}{3} & -\frac{4}{3}%
\end{array}
\right) $%
\end{tabular}
$ \\ \hline
$II$ &  \\ \hline
$F_{1R}^{l}=\left( 
\begin{array}{l}
e_{i}^{c} \\ 
\nu _{i}{}^{c} \\ 
N_{i} \\ 
N_{i}^{s}%
\end{array}
\right) _{R,\text{ }Y_{X}=\frac{1}{2}}$ & $e_{iL}^{c},$ $Y_{X}=2$ \\ \hline
$F_{1R}^{q}=\left( 
\begin{array}{l}
d_{1}^{ac} \\ 
u_{1}^{ac} \\ 
U_{1}^{ac} \\ 
T_{1}^{ac}%
\end{array}
\right) _{R,\text{ }Y_{X}=-\frac{5}{6}}$ & $%
\begin{tabular}{l}
$F_{1L}^{q}\equiv \left( 
\begin{array}{llll}
d_{1}^{a} & u_{1}^{a} & U_{1}^{a} & T_{1}^{ac}%
\end{array}
\right) _{L}$ \\ 
$\quad Y_{X}:\left( 
\begin{array}{llll}
\frac{2}{3} & -\frac{4}{3} & -\frac{4}{3} & -\frac{4}{3}%
\end{array}
\right) $%
\end{tabular}
$ \\ \hline
$F_{iL}^{q}=\left( 
\begin{array}{l}
u_{i}^{a} \\ 
d_{i}^{a} \\ 
D_{i}^{a} \\ 
H_{i}^{a}%
\end{array}
\right) _{L,\text{ }Y_{X}=-\frac{1}{6}}i=2,3$ & $%
\begin{tabular}{l}
$F_{iR}^{q}\equiv \left( 
\begin{array}{llll}
u_{i}^{a} & d_{i}^{a} & D_{i}^{a} & H_{i}^{a}%
\end{array}
\right) _{R}$ \\ 
$Y_{X}:\ \ \left( 
\begin{array}{llll}
\frac{4}{3} & -\frac{2}{3} & -\frac{2}{3} & -\frac{2}{3}%
\end{array}
\right) $%
\end{tabular}
$ \\ \hline
\end{tabular}%
\end{eqnarray*}

The second alternative is very attractive, as it can naturally accomodate
more than one right handed neutrinos per generation, some of which can be
identified with sterile neutrinos when the $SU_{L}(4)\times U_{X}(1)$
symmetry is suitably broken in the lepton sector.

We first note that $SU(4)$ lepton multiplet split as follows under the sub
group $SU_{L}(2)\times U_{Y_{1}}(1)$ of $SU(4)$ as a doublet 
\[
\left( 
\begin{array}{l}
e_{i}^{c} \\ 
\nu _{i}{}^{c}%
\end{array}
\right) _{R},\text{ }Y_{1}=\frac{1}{2}\text{ } 
\]
and two singlets 
\[
\left( 
\begin{array}{ll}
N_{i} & N_{i}^{s}%
\end{array}
\right) _{R}\text{, }Y_{1}=-\frac{1}{2}\text{ } 
\]
After breaking the group $SU(4)\times U_{X}(1)$ to $SU_{L}(2)\times U_{Y}(1)$
of the standard model, we have a doublet $\left( 
\begin{array}{l}
e_{i}^{c} \\ 
\nu _{i}^{c}%
\end{array}
\right) _{R}$ with $Y=1,$ a singlet $e_{iL}^{c}$ with $Y=2$ and two singlets 
$\left( 
\begin{array}{ll}
N_{i} & N_{i}^{s}%
\end{array}
\right) _{R}$ with $Y=0.$ It is clear that two extra neutrinos are decoupled
from the group $SU_{L}(2)\times U_{Y}(1).$

The interaction Lagrangian is given by (suppressing the generation index $i$%
) 
\begin{eqnarray}
\mathcal{L}_{I} &=&-\frac{g}{\sqrt{2}}[\overline{e}_{R}^{c}\gamma ^{\mu }\nu
_{R}^{c}W_{\mu }^{-}+h.c.]  \nonumber \\
&&-\frac{g}{\sqrt{2}}[\overline{e}_{R}^{c}\gamma ^{\mu }e_{R}^{c}(W_{3\mu }+%
\frac{g_{1}}{2g}B_{1\mu }+\frac{g_{X}}{2g}V_{\mu })+\overline{\nu }%
_{R}^{c}\gamma ^{\mu }\nu _{R}^{c}(-W_{3\mu }+\frac{g_{1}}{2g}B_{1\mu }+%
\frac{g_{X}}{2g}V_{\mu })  \nonumber \\
&&+\overline{N}_{R}\gamma ^{\mu }N_{R}(U_{3\mu }-\frac{1}{2}\frac{g_{1}}{g}%
B_{1\mu }+\frac{g_{X}}{2g}V_{\mu })+\overline{N}_{R}^{s}\gamma ^{\mu
}N_{R}^{s}(-U_{3\mu }-\frac{1}{2}\frac{g_{1}}{g}B_{1\mu }+\frac{g_{X}}{2g}%
V_{\mu })]  \nonumber \\
&&-\frac{g}{\sqrt{2}}[\overline{e}_{R}^{c}\gamma ^{\mu }N_{eR}X_{\mu }^{-}+%
\overline{\nu }_{R}^{c}\gamma ^{\mu }N_{R}X_{\mu }^{0}+\overline{e}%
_{R}^{c}\gamma ^{\mu }N_{R}^{s}Y_{\mu }^{-}+\overline{\nu }_{R}^{c}\gamma
^{\mu }N_{R}^{s}Y_{\mu }^{0}+\overline{N}_{R}\gamma ^{\mu }N_{R}^{s}U)+h.c.]
\label{13a}
\end{eqnarray}
where the vector boson $B_{1\mu }=\sqrt{\frac{2}{3}}W_{8\mu }+\sqrt{\frac{1}{%
3}}W_{15\mu }$ is coupled to $U_{Y_{1}}(1)$ and $U_{3\mu }=-\sqrt{\frac{1}{3}%
}W_{8\mu }+\sqrt{\frac{2}{3}}W_{15\mu }$. Note that in the symmetry limit $%
g_{1}=\sqrt{2}g.$ Further we note that the vector boson $B_{\mu }$
corresponding to $U_{Y}(1)$ is given by 
\begin{equation}
\frac{B_{\mu }}{g^{\prime }}=\frac{B_{1\mu }}{g_{1}}+\frac{V_{\mu }}{g_{X}}
\label{14a}
\end{equation}
Thus 
\begin{eqnarray}
A_{\mu } &=&\frac{e}{g}W_{3\mu }+\frac{e}{g^{\prime }}B_{1\mu }+\frac{e}{%
g_{X}}V_{\mu }  \nonumber \\
&=&\frac{e}{g}W_{3\mu }+\frac{e}{g^{\prime }}B_{\mu }  \nonumber \\
Z_{\mu } &=&\frac{e}{g^{\prime }}W_{3\mu }-\frac{e}{g}B_{\mu }  \label{15a}
\end{eqnarray}
There are two more vector bosons, which we define as follows 
\begin{eqnarray}
Z_{\mu }^{\prime } &=&-\frac{g_{1}}{g}B_{1\mu }+\frac{g_{X}}{g}V_{\mu } 
\nonumber \\
Z_{\mu }^{\prime \prime } &=&U_{3\mu }  \label{16a}
\end{eqnarray}
Hence rewriting the interaction Lagrangian in terms of vector bosons $A_{\mu
},$ $Z_{\mu },$ $Z_{\mu }^{\prime }$ and $Z_{\mu }^{\prime \prime }$ we have 
\begin{eqnarray}
\mathcal{L}_{I}^{neutral} &=&-g\sin \theta [\overline{e}_{R}^{c}\gamma ^{\mu
}e_{R}^{c}+\overline{e}_{L}^{c}\gamma ^{\mu }e_{L}^{c}]A_{\mu }-\frac{g}{%
2\cos \theta _{W}}[(\overline{e}_{R}^{c}\gamma ^{\mu }e_{R}^{c}-\overline{%
\nu }_{R}^{c}\gamma ^{\mu }\nu _{R}^{c})  \nonumber \\
&&-2\sin ^{2}\theta _{W}(\overline{e}_{R}^{c}\gamma ^{\mu }e_{R}^{c}+%
\overline{e}_{L}^{c}\gamma ^{\mu }e_{L}^{c})]Z_{\mu }  \nonumber \\
&&-\frac{1}{2}g[\frac{1}{2}(\overline{e}_{R}^{c}\gamma ^{\mu }e_{R}^{c}+%
\overline{\nu }_{R}^{c}\gamma ^{\mu }\nu _{R}^{c}+\overline{N}_{R}\gamma
^{\mu }N_{R}+\overline{N}_{R}^{s}\gamma ^{\mu }N_{R}^{s})  \nonumber \\
&&-\frac{g^{2}}{g_{1}^{2}}\tan ^{2}\theta _{W}(\overline{e}_{R}^{c}\gamma
^{\mu }e_{R}^{c}-2\overline{e}_{L}^{c}\gamma ^{\mu }e_{L}^{c}+\overline{\nu }%
_{R}^{c}\gamma ^{\mu }\nu _{R}^{c})]Z_{\mu }^{\prime }  \label{17a} \\
&&-\frac{1}{2}g[\overline{N}_{R}\gamma ^{\mu }N_{R}-\overline{N}%
_{R}^{s}\gamma ^{\mu }N_{R}^{s}]Z_{\mu }^{\prime \prime }  \nonumber
\end{eqnarray}
\newline

From Eqs. (\ref{13a}) and (\ref{17a}), it is clear that the doublet $\left( 
\begin{array}{l}
N_{i} \\ 
N_{i}^{s}%
\end{array}
\right) _{R}$ belongs to fundamental representation of the $U$-spin $SU(2)$
subgroup of $SU(4)$ (the other subgroup being $SU_{L}(2)$) with vector
bosons $U,\overline{U},U_{3\mu },\equiv Z_{\mu }^{\prime \prime }$ belonging
to the adjoint representation of this group. To break the symmetry and at
the same time give Dirac masses to fermions which have both left handed and
right handed components viz charged leptons and quarks, the minimally
required Higgs are given in Table 2: \pagebreak 
\begin{eqnarray*}
&& 
\begin{tabular}{|l|}
\hline
Table II \\ \hline
\end{tabular}
\text{Higgs quartets.} \\
&&\text{%
\begin{tabular}{|l|l|l|}
\hline
$Y_{X}=-\frac{3}{2}\text{ }$ &  &  \\ \hline
$\chi =\left( 
\begin{array}{l}
\chi ^{0} \\ 
\chi ^{-} \\ 
\chi ^{^{\prime }-} \\ 
\chi ^{^{\prime \prime }-}%
\end{array}
\right) $ & $<\chi >=\left( 
\begin{array}{l}
u \\ 
0 \\ 
0 \\ 
0%
\end{array}
\right) $ &  \\ \hline
$Y_{X}=+\frac{1}{2}\text{ }$ &  &  \\ \hline
$\rho =\left( 
\begin{array}{l}
\rho ^{+} \\ 
\rho ^{0} \\ 
\rho ^{^{\prime }0} \\ 
\rho ^{^{\prime \prime }0}%
\end{array}
\right) $ & $\eta =\left( 
\begin{array}{l}
\eta ^{+} \\ 
\eta ^{0} \\ 
\eta ^{^{\prime }0} \\ 
\eta ^{^{\prime \prime }0}%
\end{array}
\right) $ & $\xi =\left( 
\begin{array}{l}
\xi ^{+} \\ 
\xi ^{0} \\ 
\xi ^{^{\prime }0} \\ 
\xi ^{^{\prime \prime }0}%
\end{array}
\right) $ \\ \hline
$<\rho >=\left( 
\begin{array}{l}
0 \\ 
u^{\prime } \\ 
0 \\ 
0%
\end{array}
\right) $ & $<\eta >=\left( 
\begin{array}{l}
0 \\ 
0 \\ 
V \\ 
0%
\end{array}
\right) $ & $<\xi >=\left( 
\begin{array}{l}
0 \\ 
0 \\ 
0 \\ 
V^{^{\prime }}%
\end{array}
\right) $ \\ \hline
\end{tabular}
}
\end{eqnarray*}
The following comments are in order: $\rho $ and $\chi $ correspond to Higgs
field $\phi =\left( 
\begin{array}{l}
\phi ^{+} \\ 
\phi ^{0}%
\end{array}
\right) ,<\phi >_{0}=\left( 
\begin{array}{l}
0 \\ 
\frac{v}{\sqrt{2}}%
\end{array}
\right) ,\widetilde{\phi }=i\tau \overline{\phi }=\left( 
\begin{array}{l}
\phi _{0}^{*} \\ 
-\phi ^{-}%
\end{array}
\right) ,<\widetilde{\phi }>$ $=\left( 
\begin{array}{l}
\frac{v}{\sqrt{2}} \\ 
0%
\end{array}
\right) .$ Thus $<\rho >$ and $<\chi >$ give masses to the vector bosons of
the standard model gauge group and masses to the charged leptons and quarks
of the standard model. For charged leptons only $<\chi >$ contributes. The
other Higgs are needed to break the group $SU(4)\times U(1)$ so as to give
superheavy masses to the extra vector bosons and extra quarks outside the
standard model.

With the symmetry breaking pattern discussed above, the mass Lagrangian for
vector bosons is given by 
\begin{eqnarray}
\mathcal{L}_{mass}^{W} &=&\frac{1}{2}g^{2}u^{2}[2W^{+}W^{-}+(\frac{1}{\cos
\theta _{W}}Z+\frac{1}{2}(1-6\frac{g^{2}}{g_{1}^{2}}\tan ^{2}\theta
_{W})Z^{\prime })^{2}+2X^{+}X^{-}+2Y^{+}Y^{-}]  \nonumber \\
&&+\frac{1}{2}g^{2}u^{\prime 2}[2W^{+}W^{-}+(\frac{1}{\cos \theta _{W}}Z+%
\frac{1}{2}(1-2\frac{g^{2}}{g_{1}}\tan ^{2}\theta _{W})Z^{\prime })^{2}+2%
\overline{X}^{0}X^{0}+2\overline{Y}^{0}Y^{0}]  \nonumber \\
&&+\frac{1}{2}gV^{2}[2X^{+}X^{-}+2\overline{X}^{0}X^{0}+2U\overline{U}+(%
\frac{1}{2}Z^{\prime }+Z^{\prime \prime })^{2}]  \nonumber \\
&&+\frac{1}{2}gV^{\prime 2}[2Y^{+}Y^{-}+2\overline{Y}^{0}Y^{0}+2U\overline{U}%
+(\frac{1}{2}Z^{\prime }-Z^{\prime \prime })^{2}]  \label{18a}
\end{eqnarray}
Since $V\approx V^{\prime }\gg u\approx u^{\prime },$ therefore neglecting
the terms of order $\frac{u^{2}}{V^{2}},$ we can write 
\begin{eqnarray}
\mathcal{L}_{mass}^{W} &\approx &\frac{1}{2}g^{2}[(u^{2}+u^{\prime
2})(2W^{+}W^{-}+\frac{1}{\cos ^{2}\theta _{W}}Z^{2})]+\frac{1}{2}%
g^{2}[V^{2}(2X^{+}X^{-}+2X^{0}\overline{X}^{0}+2U\overline{U})  \nonumber \\
&&+(\frac{1}{2}Z^{\prime }+Z^{\prime \prime })^{2})+V^{\prime 2}((\frac{1}{2}%
Z^{\prime }-Z^{\prime \prime })^{2}+2Y^{+}Y^{-}+2Y^{0}\overline{Y}^{0}+2U%
\overline{U})]  \label{19a}
\end{eqnarray}
This gives the gauge boson masses 
\begin{eqnarray*}
m_{W}^{2} &=&\frac{1}{2}g^{2}(u^{2}+u^{\prime 2})=\frac{m_{Z}^{2}}{\cos
^{2}\theta _{W}} \\
m_{X}^{2} &=&\frac{1}{2}g^{2}V^{2},\text{ }m_{Y}^{2}=\frac{1}{2}%
g^{2}V^{\prime 2},\text{ }m_{U}^{2}=m_{X}^{2}+m_{Y}^{2} \\
m_{Z^{\prime }Z^{\prime \prime }}^{2} &=&\frac{1}{2}g^{2}\left( 
\begin{array}{ll}
\frac{V^{2}+V^{\prime 2}}{4} & \frac{V^{2}-V^{\prime 2}}{2} \\ 
\frac{V^{2}-V^{\prime 2}}{2} & V^{2}+V^{\prime 2}%
\end{array}
\right)
\end{eqnarray*}
So far we have introduced essentially two energy scales, represented by the
vacuum expectation values $u$($\simeq u^{\prime }$) and $V$($\simeq
V^{\prime }$) the former corrsponds to the SM scale ($\simeq 175$GeV) and
the latter although not fixed, but much higher, an intersting one would be
an intermediate energy scale (between SM and GUT)$\simeq 10^{10}$GeV.

Lagrangian (\ref{17a}) explicitly shows the decoupling of extra leptons from
the standard model. Only connection with the standard model is through the
mixing of $Z$ with $Z^{\prime }$ (involving terms of order $\frac{u^{2}}{%
V^{2}})$ when the symmetry is broken to give masses to the gauge bosons.
Since extra bosons other than the standard bosons are very heavy, the mixing
term only give a negligible contribution to the standard model observables.
Similarly $Z^{\prime }$ being very heavy, its contribution to standard model
observables is also negligible.

We now discuss $SU_{L}(4)\times U_{X}(1)$ Yukawa interaction in the charged
lepton sector, which is 
\[
H_{Y}=h_{ij}(\overline{F}_{iR}^{c}\chi _{j}e_{iL}^{c})+h.c. 
\]
so that 
\begin{equation}
M_{l}=h_{ij}(\overline{e}_{iL}u_{j}e_{iR})+h.c.  \label{17}
\end{equation}
For simplicity we may take $u_{e}=u_{\mu }=u_{\tau }=u.$ It is important to
note that we have no new charged leptons, other than those of the standard
model. As far as new quarks are concerned, their masses will be determined
essentially by $h_{iQ}V$, $h_{iQ}$ being the corresponding Yukawa coupling
constant, Q stands for U, T, D and H quarks. If Yukawa couplings are of
order unity, masses of new quarks will be of same order as those of vector
bosons X etc.

The above scalars do not give masses to neutrinos. In a way this is a great
advantage as neutrinos have a completely different mass spectrum from other
fermions.

To give Majorana masses to neutrinos, we introduce Higgs scalars $S_{\alpha
\beta }$ $[\alpha ,\beta $ are $SU(4)$ indices$]$ belonging to the symmetric
representation $10$ of $SU(4)$ with $Y_{X}=1.$ The electric charge matrix
for this representatin is 
\[
\widehat{Q}=\left( 
\begin{array}{llll}
2 & 1 & 1 & 1 \\ 
1 & 0 & 0 & 0 \\ 
1 & 0 & 0 & 0 \\ 
1 & 0 & 0 & 0%
\end{array}
\right) 
\]
Thus charged leptons can not get any masses from the Higgs $S_{\alpha \beta
} $ nor the quarks. In order to make $N$ heavy and $N^{s}$ light ($\sim $eV)
we introduce 
\[
\overline{S}_{\alpha \beta },\text{ \qquad }\overline{S}_{\alpha \beta
}^{^{\prime }}\text{, and }\overline{S}_{\alpha \beta }^{^{\prime \prime }}%
\text{with }Y_{X}=-1 
\]
with their expectation values in Table 3. 
\begin{eqnarray*}
&& 
\begin{tabular}{|l|}
\hline
Table III \\ \hline
\end{tabular}
\text{Vacuum expectation values of Higgs belonging to rep. 10 of SU(4)} \\
&& 
\begin{tabular}{|l|l|}
\hline
$<\overline{S}_{\alpha \beta }^{\prime }\text{ }>=\left( 
\begin{array}{llll}
0 & 0 & 0 & 0 \\ 
0 & 0 & \kappa ^{\prime } & 0 \\ 
0 & \kappa ^{\prime } & 0 & 0 \\ 
0 & 0 & 0 & 0%
\end{array}
\right) $ & $<\overline{S}_{\alpha \beta }^{^{\prime \prime }}\text{ }%
>=\left( 
\begin{array}{llll}
0 & 0 & 0 & 0 \\ 
0 & 0 & 0 & \kappa _{s}^{\prime } \\ 
0 & 0 & 0 & 0 \\ 
0 & \kappa _{s}^{\prime } & 0 & \kappa _{s}%
\end{array}
\right) $ \\ \hline
$<\overline{S}_{\alpha \beta }\text{ }>=\left( 
\begin{array}{llll}
0 & 0 & 0 & 0 \\ 
0 & 0 & 0 & 0 \\ 
0 & 0 & \kappa _{R} & 0 \\ 
0 & 0 & 0 & 0%
\end{array}
\right) $ &  \\ \hline
\end{tabular}%
\end{eqnarray*}
where $\kappa _{s},\kappa _{s}^{\prime }\ll \kappa ^{\prime }\ll \kappa
_{R}. $ We wish to remark that we have selected three different Higgs
scalars $S_{\beta \alpha }$ with the same $Y_{X}=-1$ , just as we selected
three Higgs quartets $\rho ,$ $\eta ,$ $\xi $ with $Y_{X}=-\frac{1}{2},$ for
the reason that with one $S_{\beta \alpha }$ the vacuum expectation values $%
\kappa ^{\prime },\kappa _{R},\kappa _{s},\kappa _{s}^{\prime }$ as
different components of the same Higgs scalar would have been critically
alligned.They, now belonging to three different Higgs scalars, are of course
very hierarchical to accomodate different energy scales. However, the
problem of hierarchy is there for models which go beyond the standard model
and has no easy solution as is well known. These additional Higgs give an
extra contribution to $\mathcal{L}_{mass}^{W}$ in Eq. (\ref{19a}) [only $%
\kappa _{R}$ is important] 
\[
\frac{1}{2}g^{2}\kappa _{R}^{2}[X^{+}X^{-}+\overline{X}^{0}X^{0}+2(\frac{1}{2%
}Z^{\prime }+Z^{\prime \prime })^{2} 
\]
so that only X and Z$^{\prime }$and Z$^{\prime \prime }$ gauge bosons get
extra contribution, giving further splitting between these and Y and U
bosons.

The Yukawa couplings of these scalars to leptons are 
\begin{equation}
\sum_{b}f_{ij}^{b}F_{i\alpha }^{T}C^{-1}F_{j\beta }\overline{S}_{\beta
\alpha }^{b}  \label{19}
\end{equation}
where $b=$ no prime, prime and double prime. Then the neutrino mass matrix
is given by 
\begin{equation}
M_{N}=f_{ij}(\kappa _{R}N_{i}^{T}C^{-1}N_{j})+f_{ij}^{\prime }\kappa
^{\prime }\overline{\nu }_{i}N_{j}+f_{ij}^{\prime \prime }(\kappa
_{s}N_{i}^{sT}C^{-1}N_{j}^{s}+\kappa _{s}^{\prime }\overline{\nu }%
_{i}N_{j}^{s})+h.c  \label{20}
\end{equation}
Eq. (\ref{20}) gives 9$\times $9 neutrino mass matrix in the weak
eigenstates basis: 
\begin{equation}
M_{\nu _{N}}=\left( 
\begin{array}{lll}
0 & m_{D} & m_{D}^{s} \\ 
m_{D}^{T} & M & 0 \\ 
m_{D}^{_{s}T} & 0 & m^{s}%
\end{array}
\right)  \label{21}
\end{equation}
where ($m_{D}$)$_{ij}=f_{ij}^{\prime }\kappa ^{\prime },$ ($M$)$%
_{ij}=f_{ij}\kappa _{R},$($m_{D}^{s}$)$_{ij}=f_{ij}^{\prime \prime }\kappa
_{s}^{\prime },$ ($m^{s}$)$_{ij}=f_{ij}^{\prime \prime }\kappa _{s}$.

Since $\kappa _{s},$ $\kappa _{s}^{\prime }<<\kappa ^{\prime },\ll \kappa
_{R},$ therefore in the limit $m_{D}^{s},m^{s}\rightarrow 0,$ the
diagonalization gives 
\begin{equation}
M_{\nu _{N}}=\left( 
\begin{array}{lll}
A & 0 & 0 \\ 
0 & M & 0 \\ 
0 & 0 & 0%
\end{array}
\right)
\end{equation}
where $A=-m_{D}M^{-1}m_{D}^{T}$. However, by introducing a unitary matrix U: 
\begin{eqnarray}
U &=&\left( 
\begin{array}{lll}
1 & b & 0 \\ 
-b^{T} & 1 & 0 \\ 
0 & 0 & 1%
\end{array}
\right) ,\text{ }  \nonumber \\
b &=&-m_{D}M^{-1},b^{T}=-M^{-1}m_{D}^{T}
\end{eqnarray}
we obtain the mass matrix 
\begin{eqnarray}
UM_{\nu _{N}}U^{T} &=&\left( 
\begin{array}{lll}
A & 0 & m_{D}^{s} \\ 
0 & M & 0 \\ 
m_{D}^{_{s}T} & 0 & m_{s}%
\end{array}
\right) +O(\frac{1}{M^{2}})  \nonumber \\
&=&M\left( 
\begin{array}{lll}
0 & 0 & 0 \\ 
0 & 1 & 0 \\ 
0 & 0 & 0%
\end{array}
\right) +\left( 
\begin{array}{lll}
A & 0 & 0 \\ 
0 & 0 & 0 \\ 
0 & 0 & 0%
\end{array}
\right) +\left( 
\begin{array}{lll}
0 & 0 & m_{D}^{s} \\ 
0 & 0 & 0 \\ 
m_{D}^{_{s}T} & 0 & m_{s}%
\end{array}
\right) +O(\frac{1}{M^{2}})
\end{eqnarray}
This gives the effective $6\times 6$ light neutrino mass matrix 
\begin{equation}
M_{\nu }=\left( 
\begin{array}{ll}
-m_{D}M^{-1}m_{D}^{T} & m_{D}^{s} \\ 
m_{D}^{_{s}T} & m_{s}%
\end{array}
\right)  \label{26a}
\end{equation}
The phenomenology of this $6\times 6$ matrix already exist in the literature
[4,5,12,13]. In Eq. (\ref{26a}) $-m_{D}M^{-1}m_{D}^{T}$ gives the normal see
saw mechanism. Here $M>>m_{D}$ and may be of order $10^{10}-10^{12}$GeV a
scale which may be relevant for thermal, non-degenerate leptogenesis. Now a
word about energy scales: $\kappa _{R}$ is of order of 10$^{10}$GeV while $%
\kappa ^{\prime }$ is of order 1GeV so that if all Yukawa coupling constants
in Eq.(\ref{20}) are of order unity, active neutrinos mass is of order 0.1eV
to satisfy the constraint from WMAP data $\Sigma m_{\nu _{i}}<(0.4-0.7)$eV
while atmospheric data give $m_{\nu }>5\times 10^{-2}$eV.$\kappa _{s}$ is of
order a few eV so that sterile neutrinos are to be relevant for ``short''
baseline oscillation searches. Finally in order to have active-sterile
mixing small($\sim 0.1$)[see ref 11, 12]$\kappa _{s}^{\prime }\simeq
0.1\kappa _{s}.$ For other approaches for possible existance of sterile
neutrinos in eV scale range whithin the framework of higher order gauge
groups beyond the standard model., see ref [13,14,15].

In summary we have shown (i) intermediate mass scales between electroweak
mass scale and that of grand unification can be accommodated, relevent for
leptogeneses (ii) one can naturally accommodate more than one right handed
neutrino per generation, some of which can be identified with sterile
neutrinos that mix with active ones (iii) neutrinos mass spectrum is
decoupled from other fermions (iv) both the normal seesaw mechanism with
right handed neutrino Majorana mass $M\gg M_{\text{weak}}$, which may be
relevant for leptogenesis and an eV-scale which may manifest itself in
``short'' baseline oscillation searches neutrino experiments.

One of the authors(R) would like to thank Prof. K Sreenivasan for
hospitality at Abdus Salam International Centre for Theoretical Physics,
where most of this work was done. The other author (F) acknowledges a
research grant provided by the Higher Education Commission of Pakistan to
him as a Distinguished National Professor.

\end{document}